\documentclass[preprint,showpacs,preprintnumbers,amsmath,amssymb]{revtex4}
\usepackage{graphicx}% Include figure files
\usepackage{dcolumn}% Align table columns on decimal point
\usepackage{bm}% bold math
\usepackage{portland}  % landscape
\usepackage{rotating}  % rotate materials
\begin{document}
\title{Correlations among centrality measures in complex networks}
\author{Chang-Yong Lee}
\email{clee@kongju.ac.kr}
\affiliation{The Department of Industrial Information, Kongju National
  University, Chungnam, 340-702 South Korea}
\date{\today}

\begin{abstract}
In this paper, we empirically investigate correlations among four
centrality measures, originated from the social science, of various
complex networks. For each network, we compute the centrality
measures, from which the partial correlation as well as the
correlation coefficient among measures is estimated. We uncover that
the degree and the betweenness centrality are highly correlated;
furthermore, the betweenness follows a power-law
distribution irrespective of the type of networks. This characteristic
is further examined in terms of the conditional probability
distribution of the betweenness, given the degree. The conditional
distribution also exhibits a power-law 
behavior independent of the degree which explains partially, if not
whole, the origin of the power-law distribution of the betweenness. A
similar analysis on the random network reveals that these 
characteristics are not found in the random network.
\end{abstract}
\maketitle
%%%%%%%%%%%%%%%%%%%%%%%%%%%%%%%%%%%%%%%%%%%%%%%%%%%%%
\section{Introduction}
%%%%%%%%%%%%%%%%%%%%%%%%%%%%%%%%%%%%%%%%%%%%%%%%%%%%%
The network (or graph) is a useful way of expressing and 
investigating quantitatively the characteristics of complex systems in
various disciplines. It consists of a set of vertices representing
entities, and edges associated with connections between
vertices~\cite{review}. Numerous complex systems can be and have been 
expressed in terms of networks, and they are often classified by the
research field, such as social~\cite{amaral,newmansocial},
technological~\cite{fal,albertweb}, and biological
networks~\cite{guelzin,jeong}, to name just a few. 

Early researches on the network focused mainly on the regular and
the random networks from which many mathematical results for
general structural characteristics have been extracted~\cite{er}. 
Recently, due to the availability of computers and the Internet, study
on large-scale statistical properties of complex networks has been
possible. It was found that many complex networks had distinctive
features in common, such as the power-law distribution of the
degree and the clique of the network, resulting in the
scale-free~\cite{scale} and the small world
networks~\cite{watts}. These uncovered characteristics, which differ
from those of the regular and random networks, was the trigger that
brought about considerable advances in the understanding of complex
networks, including the development of numerous analysis tools and
devising more accurate topological models for the observed
networks~\cite{review}.  

More recently, the research on complex networks drifts also toward 
the community structure of the networks. It has been shown that various
complex networks can be organized in terms of the community
structure (or modularity), in which groups of vertices that are highly
interconnected have looser connections between them. 
The analysis of these structures has been a topic of intensive investigation
in conjunction with many practical relevance, such as finding functional
modules in biological networks~\cite{ravasz,rives} and identifying
communities in the Web in order to construct, for instance, an
efficient search engine~\cite{flake}. 

Various attempts have been made to find or identify community
structures in complex networks~\cite{creview}. Examples include the
hierarchical clustering~\cite{wasserman}; methods based on the edge
betweenness~\cite{girvan,tyler}, the edge-clustering via the 
degree~\cite{radicchi}, the information centrality~\cite{clauset}, and
the eigenvector centrality~\cite{yang}; the information-theoretic
approach via the degree~\cite{ziv}. These methods are directly
or indirectly related to the centrality measures. 
Considering that resulting community structure depends on
the choice of measures (including the centrality measures) adopted in
various schemes, it would be interesting to investigate any relation
among the centrality measures.  

The centrality (or sociometric status) has been studied
particularly in the social science from the perspective of the 
social connectivity. It is an
incarnations of a concept that describes vertices' prominence and/or
importance in terms of features of their network
environment~\cite{friedkin}. It addresses an issue of which 
individuals are best connected to other or have most influence. This
relative importance was quantified by various measures, developed
mainly by researchers of the social networks~\cite{freeman}. Different
measures for the centrality have been proposed in the social
science. Among them, four centrality measures are commonly used in
the network analysis: the degree, the closeness, the betweenness, and
the eigenvector centrality~\cite{friedkin,bonacich,freeman}.  

In this paper, we empirically investigate correlations among the centrality 
measures in complex networks to gain some insight into the potential role of
the measures in analyzing complex networks. We restrict our analysis to
undirected networks, since some of centrality measures, such as the
eigenvector centrality, cannot be defined unambiguously for directed networks. 
We analyze the film actor network, the scientific collaboration
network, the neural network of {\it Caenorhabditis elegans}, the
Internet of both the Autonomous System (AS) and the router levels, and
protein interaction networks. Analyzed organisms for protein
interaction networks are {\it Saccharomyces cerevisiae}, {\it
  Escherichia coli}, {\it Caenorhabditis elegans}, {\it Drosophila
  melanogaster}, {\it Helicobacter pylori}, and {\it Homo
  sapiens}~\cite{data}.   
%%%%%%%%%%%%%%%%%%%%%%%%%%%%%%%%%%%%%%%%%%%%%%%%%%%%
\section{Centrality Measures}
%%%%%%%%%%%%%%%%%%%%%%%%%%%%%%%%%%%%%%%%%%%%%%%%%%%%%
The centrality measures are introduced as a way of specifying and 
quantifying the centrality concept of a vertex in a
network. Furthermore, they are often classified according to the
extent to which a vertex has influence on the others: the immediate
effects, the mediative effects, and the total effects
centrality~\cite{friedkin}. Typical examples which 
belong to each class are: the closeness and degree for the immediate;
the betweenness for the mediative; the eigenvector for the total effect
centrality. In addition, these measures are argued to be 
complementary rather than competitive because they stem from the same
theoretical foundation~\cite{friedkin}. Although the measures are
well known, we restate them here for the completeness with the
emphasis on their implications.

The degree centrality is the most basic of all measures which counts
how many vertices are involved in an interaction. It is defined, for a
vertex $i$, as the number of edges that the vertex has. That is,
\begin{equation}
d_{i}=\sum_{j=1}^{n} a_{ij}~,
\end{equation}
where $n$ is the number of vertices in the network, and $a_{ij}=1$ if
vertices $i$ and $j$ are connected by an edge, $a_{ij}=0$
otherwise. It measures the opportunity to receive information flowing
through the network with everything else being equal. The degree is
also a prominent quantity whose distribution follows a power-law
distribution in scale-free networks~\cite{scale}. 

The eigenvector centrality can be understood as a
refined version of the degree centrality in the sense that it
recursively takes into account how neighbor vertices are connected. That
is, the eigenvector centrality $e_{i}$ of a vertex $i$ is proportional
to the sum of the eigenvector centrality of the vertices it is
connected to. It is defined as 
\begin{equation}
e_{i}=\lambda^{-1} \sum_{j} a_{ij}~ e_{j}~,
\end{equation}
where $\lambda$ is the largest eigenvalue to assure the centrality is
non-negative. Thus, $e_{i}$ is the $i$th component of the
eigenvector associated with the largest eigenvalue $\lambda$ of the
network. While the eigenvector centrality of a network can be
calculated via the standard method~\cite{recipes} using the adjacent 
matrix representation of the network, it can be also 
computed by an iterative degree calculation, known as the accelerated power
method~\cite{hotelling}. This method is not only more efficient, but
consistent with the spirit of the refined version of the degree centrality. 

The closeness centrality stems from the notion that the influence of
central vertices spreads more rapidly throughout a network than that
of peripheral ones. It is defined, for each vertex $i$, as 
\begin{equation}
c_{i}=\left( \sum_{j}~d_{ij} \right)^{-1}~,
\end{equation}
where $d_{ij}$ is the length of the shortest path (geodesic)
connecting vertices $i$ and $j$. Thus, the closeness is closely associated
with the characteristic path length~\cite{watts}, the average path
length of all paths between all pairs of vertices.

The betweenness centrality, or the load~\cite{load}, is a measure of
the influence of a vertex over the flow of information between every
pair of vertices under the assumption that information primarily flows
over the shortest path between them. It measures the accumulated
number of information transmissions that occur through
the pass. The removal of high betweenness vertices sometimes results
in disconnecting a network. The betweenness centrality of a vertex $i$
is defined as
\begin{equation}
b_{i}=\sum_{jk}^{n}\frac{g_{jk(i)}}{g_{jk}}~,
\end{equation}
where $g_{jk}$ is the number of geodesics between $j$ and $k$, and
$g_{jk(i)}$ is the number of geodesics that pass through $i$ among
$g_{jk}$. Since $b_{i}$ is of the order ${\cal O}(n^{2})$,
in this paper, we normalize $b_{i}$ with its maximum value of
$(n-1)(n-2)/2$ so that $b_{i} \in [0, 1]$ for all $i$. 
%%%%%%%%%%%%%%%%%%%%%%%%%%%%%%%%%%%%%%%%%%%%%%%%%%%%%%%%%%%%
\section{Correlation Analysis}
%%%%%%%%%%%%%%%%%%%%%%%%%%%%%%%%%%%%%%%%%%%%%%%%%%%%%%%%%%%%
\subsection{Correlation coefficients and partial correlations}
%%%%%%%%%%%%%%%%%%%%%%%%%%%%%%%%%%%%%%%%%%%%%%%%%%%%%%%%%%%%%%%%%%%%
For every network, we compute the four centrality measures so that
all four values are assigned to each vertex. The correlation between
a pair of different measures can be estimated by the correlation
coefficient~\cite{chao}. More specifically, it is a quantity which
measures the linear correlation between vertex-wise pairs of data,
$(A, B)=\{(a_{i}, b_{i}), ~i=1, 2, \cdots , n\}$, and is given as
\begin{equation}
R_{AB}=\frac{\sum \left( a_{i}- \bar{A} \right) \left(
    b_{i}- \bar{B} \right)}{n~ \sigma_{A} ~\sigma_{B}}~,
\end{equation}
where $\bar{A}$ and $\sigma_{A}$ are the mean and standard deviation
of the measurements of a centrality measure $A$. The value of $R_{AB}$
ranges from -1 to 1: 1 being totally correlated, and -1 being totally
anti-correlated.  

Table I shows correlation coefficients estimated between
pairs of data obtained from different centrality measures. As shown in
Table I, the degree is strongly correlated with the
betweenness and less strongly with the eigenvector centrality;
whereas the closeness is weakly correlated with the other
measures. This implies that the three measures (the degree,
the betweenness, the eigenvector centrality) are closely
inter-related. In general, correlation coefficients estimated
from different variables could be significantly
overlapped. That is, a certain amount of correlation found 
between any two measures may be tied in with correlations with the third.  

To take into account this point, we introduce the partial correlation
method~\cite{johnston}. The partial correlation is a method that
determines the correlation between any two variables under the
assumption that each of them is not correlated with the third. That
is, it estimates the correlation between two variables while the third
variable is held constant. Formally, the partial correlation between
variables $A$ and $B$ while holding $C$ constant is given in terms of
the corresponding correlation coefficients as 
\begin{equation}
R_{AB \cdot C}=\frac{R_{AB}- R_{BC}~ R_{AC}}{\sqrt{ \left(1-R^{2}_{BC}
    \right) \left(1-R^{2}_{AC} \right)}}~.
\end{equation}
We estimate all possible partial correlations for each correlation
coefficient, and results are shown in the parentheses of Table I.

From Table I, we find that the partial correlation between the degree
and the betweenness, while holding either the eigenvector or the
closeness constant, differs little from the correlation
coefficient between them. This implies that the strong correlation between the
degree and the betweenness is solely due to the two measures by 
themselves, and little affected by other measures. 
In contrast, the partial correlation between the betweenness and the
eigenvector (or the betweenness and the closeness) while holding the
degree constant is anti-correlated. This implies that the
positive correlation between the betweenness and the eigenvector (or
the betweenness and the closeness) is almost entirely due to
correlations with the degree. That is, a {\it positive}
correlation between them would change dramatically to a {\it negative}
correlation if they were not correlated with the degree centrality. 
Table I also shows that the correlation between the degree and the
eigenvector is affected by the betweenness and closeness. 
%%%%%%%%%%%%%%%%%%%%%%%%%%%%%%%%%%%%%%%%%%%%%%%%%%%%%%%%%%%%
\subsection{Probability distribution of the betweenness}
%%%%%%%%%%%%%%%%%%%%%%%%%%%%%%%%%%%%%%%%%%%%%%%%%%%%%%%%%%%%
From the correlation analysis, we uncover that the degree and the
betweenness are correlated much strongly than other centrality
measures. This is, in a sense, expected since vertices of a high
degree would have better chance to be included in the shortest path along
a pair of vertices. To address the correlation between
the degree $k$ and the betweenness $b$, we relate them, 
via the Bayes' theorem, as 
\begin{equation}
P(b)=\sum_{k} P(b|k)~ P(k)~,
\label{condi}
\end{equation}
and focus on the conditional probability distribution $P(b|k)$ of $b$
given $k$. To obtain reliable statistics for the conditional
distribution, we choose the film actor network as an
example since it is composed of the largest number of vertices (over
370,000 vertices) in this study. Figure 1(a) shows a few conditional 
probability distributions $P(b|k)$. As shown in
Fig.~1(a), the conditional distribution approximately follows a
power-law form with its exponent $f(k)$ depending on $k$, i.e.,
\begin{equation}
P(b|k) \propto b^{-f(k)} ~.
\label{power}
\end{equation}
The $k$-dependent exponent $f(k)$ can be estimated from 
different degrees $k$. As Fig.~1(b) suggests, $f(k)$ depends
roughly linearly on $k$. Thus, we have
\begin{equation}
f(k) \approx \alpha k + \beta ~,
\label{linear}
\end{equation}
where parameters $\alpha$ and $\beta$ can be estimated by the least
square fit. 

With Eq.~(\ref{power}) and (\ref{linear}), the probability distribution
$P(b)$ of the betweenness $b$ can be expressed as
\begin{equation}
P(b)  \propto    b^{- \beta} \sum_{k} b^{-\alpha k} P(k)~.
\end{equation}
Under the assumption that $P(k)$ does not blow up as $k$ increase, the
dominant contribution of the summation comes from small values of $k$. 
Thus, to the first approximation, we find that the betweenness follows
a power-law distribution, independent of the degree distribution. That is,
\begin{equation}
P(b) \propto b^{-(\alpha+\beta)}~,
\end{equation}
with $\alpha+\beta=2.89$ for the film actor network. 

The power-law distribution of the betweenness can also be obtained by
the direct estimate of the betweenness distribution. Figure 2 shows
betweenness probability distributions of a few networks.
Scale-free networks, such as the film actor and the protein
interaction network of
{\it D. melanogaster}, have a power-law in the distribution of the
betweenness which was first found in Ref.~\cite{load}. Considering
that the degree is highly correlated with the betweenness, it is not
surprising that the betweenness of scale-free networks follows a
power-law distribution. From Fig.~2, we also find that the directly estimated
exponent 2.36 for the film actor network is close to the derived
exponent of $\alpha+\beta=2.89$. 

Moreover, Fig.~2 shows that the power-law distribution of the
betweenness is not restricted to the scale-free network, but held true
to other types of networks, such as the collaboration network and the
neural network of {\it C. elegans}. Furthermore, as depicted in Fig.~3, the
conditional probability distribution of non scale-free networks, 
for instance, the collaboration network, is also approximately a power-law 
distribution; furthermore, the exponent of the distribution is
insensitive to the degree $k$.  

The power-law of the
conditional probability distribution is less clear for networks of
small number of vertices. This is probably due to insufficient number
of data to obtain reliable statistics. We, however, have seen the
power-law of the conditional distribution for networks 
composed of relatively sufficient number of vertices, irrespective of
the type of networks. From this, we may infer that it is the power-law
of the conditional probability distribution that is
responsible for the power-law nature of the betweenness. 

For a comparison, we apply the same analysis as above to the random
network. Table II shows correlation coefficients and partial
correlations between measures estimated for the random network. In
contrast to the real networks, every centrality measure is very
strongly correlated with every other measures. This distinctive
characteristic, however, changes dramatically once we
introduce the partial correlation. From partial correlation estimates,
we find that correlation coefficients between all possible pairs of 
measures, except that between the degree and the
betweenness, contain considerable amount of correlation tied in with
the other measures. Similar to the real networks, a strong correlation
between the degree and the betweenness is nearly maintained when these
measures are assumed not to be correlated with the other measures. 

We also examine the conditional probability distribution of the
betweenness given the degree. A few conditional distributions $P(b|k)$
of the betweenness $b$ given the degree $k$ are depicted in
Fig.~4. Unlike complex networks, the distribution is not a power-law,
but approximately a Gaussian irrespective of the degree. Since the
conditional distribution of the betweenness given the degree does not
follow a power-law distribution, we expect that the betweenness
distribution of the random network may as well differ from that of
real networks. As shown in Fig.~5, it turns out that the betweenness
distribution for the random network can be approximated as a
log-normal distribution, 
\begin{equation}
P(b)=\frac{1}{\sqrt{2\pi}~ \sigma b}~ e^{-(\ln b-\mu)^{2}/ 2 \sigma^{2}}~,
\end{equation}
where $\mu$ and $\sigma$ are the scale and the shape
parameters of the distribution, respectively. 
%%%%%%%%%%%%%%%%%%%%%%%%%%%%%%%%%%%%%%%%%%%%%%%%%%%%%%%%%%%%
\section{Summary and Conclusion}
%%%%%%%%%%%%%%%%%%%%%%%%%%%%%%%%%%%%%%%%%%%%%%%%%%%%%%%%%%%%
In this paper, in order to investigate correlations among the
measures, we applied four centrality measures (the degree,
the closeness, the betweenness, and the eigenvector centrality) to
various types of complex networks as well as the random
network. We found that the degree was strongly correlated with the
betweenness, and the correlation was robust in the sense
that the extent of correlation was little affected 
by the presence of the other measures. This finding was confirmed by
estimating the partial correlation between the degree and the
betweenness, while holding either the eigenvector or the closeness
constant. 

Based on the strong correlation existed between the two measures, we
further uncovered the  
characteristics of the betweenness. Not only for scale-free networks
but for other types of networks, the conditional distribution of the
betweenness given the degree was approximately a power-law  which, in
turn, played a predominant role in understanding the power-law
distribution of the betweenness. This feature was distinct 
from the random networks in which the conditional distribution was
roughly a Gaussian.

Within complex networks, the scale-free network by itself implies the
existence of a hierarchy with respect to the degree
centrality~\cite{ravasz2,ravasz}. Similarly, the power-law
distribution of the betweenness may suggest a new potential role of the
betweenness in quantifying the hierarchy in conjunction with the
community structure~\cite{creview}. Therefore, it may 
provide us with feasibility to use the betweenness and/or related
quantities as a measure for constructing hierarchical and community
structures of complex networks. 
%%%%%%%%%%%%%%%%%%%%%%%%%%%%%%%%%%%%%%%%%%%%%%%%%%%%%%%%%%%%
\begin{acknowledgments}
%%%%%%%%%%%%%%%%%%%%%%%%%%%%%%%%%%%%%%%%%%%%%%%%%%%%%%%%%%%%
We like to thank M. Newman for providing us with the scientific
collaboration network data. We also appreciate the open sources of various
complex network data available at many URLs.
This work was supported by the Korea Research Foundation Grant funded
by the Korean Government (MOEHRD) (KRF-2005-041-H00052). 
\end{acknowledgments}
%%%%%%%%%%%%%%%%%%%%%%%%%%%%%%%%%%%%%%%%%%%%%%%%%%%%%%%%%%%%%%%%%%%%%

%%%%%%%%%%%%%%%%%%%%%%%%%%%%%%%%%%%%%%%%%%%%%%%%%%%%%%%%%%%%
\newpage
%%%%%%%%%%%%%%%%%%%%%%%%%%%%%%%%%%%%%%%%%%%%%%%%%%%%%%%%%%%%
%%%%%%%%%%%%%%%%%%%%%%%%%%%%%%%%%%%%%%%%%%%%%%%%%%%%%%%%%%%%
\begin{table*}
\begin{ruledtabular}
\caption{Correlation coefficients and corresponding partial
  correlations (in the parentheses) between pairs of centrality
  measures for each network. $X$ stands for the degree centrality;
  while $Y$, $Z$, and 
  $W$ stand for the betweenness, the eigenvector, and the closeness
  centrality, respectively. Note that the notation for the partial
  correlation is abbreviated in such a way that corresponding two
  variables are replaced by a ``big dot''.} 
\begin{tabular}{l c c c c c c}
 & $R_{XY}$  & $R_{XZ}$ & $R_{YZ}$ & $R_{XW}$  & $R_{YW}$ & $R_{ZW}$
 \\ [-1.0ex]
\raisebox{1.5ex}[0pt]{Network}  
  & ($R_{\bullet Z}, R_{\bullet W}$) & ($R_{\bullet Y}, R_{\bullet
  W}$)  & ($R_{\bullet Y}, R_{\bullet Z}$) & ($R_{\bullet X},
  R_{\bullet Z}$) & ($R_{\bullet X}, R_{\bullet Y}$) & ($R_{\bullet
  X}, R_{\bullet W}$) \\  \hline
& 0.81 & 0.61 & 0.26 & 0.31 & 0.20 & 0.22 \\ [-1.0ex]
\raisebox{1.5ex}[0pt]{Film actor}
&  (0.85, 0.81)  &  (0.71, 0.59)   &  (-0.50, 0.23)
&  (0.27, ~0.23) &  (-0.10, ~0.15) &  (0.04, 0.18) \\ 
&  0.98 & 0.82 & 0.79 & 0.19 & 0.16 & 0.60 \\ [-1.0ex]
\raisebox{1.5ex}[0pt]{Internet (AS)} 
& (0.94, 0.98) &  (0.38, 0.91) & (-0.12, 0.88)
& (0.16, -0.68) & (-0.12, -0.65) & (0.80, 0.79) \\ 
& 0.58 & 0.36 & 0.23 & 0.29 & 0.15 & 0.12 \\ [-1.0ex]
\raisebox{1.5ex}[0pt]{Internet (router)}
&  (0.55, 0.57) & (0.28, 0.34) & ( 0.03, 0.21) &  (0.26, ~0.27) &
(-0.03, ~0.13) &   (0.02, 0.09)  \\ [1.0ex]
&  0.72 &  0.53 & 0.26  & 0.56   &  0.40  & 0.33  \\ [-1.0ex]
\raisebox{1.5ex}[0pt]{Collaboration}
  &  (0.71, 0.65) &  (0.52, 0.45) &  (-0.21, 0.14)
&  (0.43, ~0.49) & (-0.00, ~0.35) &  (0.04, 0.25)   \\
& 0.73  & 0.95  & 0.58  & 0.90  & 0.58  &  0.91  \\ [-1.0ex]
\raisebox{1.5ex}[0pt]{Neural network}
  &  (0.70, 0.59) &  (0.95, 0.74) &  (-0.53, 0.17)
&  (0.86,  ~0.29) &  (-0.26, ~0.15) &  (0.37, 0.86)  \\
&  0.88 & 0.82  & 0.62  & 0.57  & 0.34   &  0.68 \\ [-1.0ex]
\raisebox{1.5ex}[0pt]{{\it S. cerevisiae}}
 &  (0.83, 0.89) &  (0.74, 0.72) &  (-0.38, 0.57) &  (0.59, ~0.02) &
 (-0.40, -0.14) &  (0.45, 0.63)  \\  
&  0.82  & 0.75  &  0.57  & 0.20  & 0.18  & 0.68  \\ [-1.0ex]
\raisebox{1.5ex}[0pt]{{\it E. coli}}
  & (0.73, 0.82) &  (0.60, 0.86) & (-0.12, 0.62)
&  (0.08, -0.65) &  (~0.04, -0.34) &  (0.82, 0.72)  \\ 
& 0.96  &  0.74  & 0.71 &  0.41  & 0.37  &  0.60  \\ [-1.0ex]
\raisebox{1.5ex}[0pt]{{\it C. elegans}}
 &  (0.92, 0.95) & (0.32, 0.68) & (-0.03, 0.65) & (0.22, -0.05) &
 (-0.10, -0.09) &  (0.47, 0.51)  \\  
& 0.91   & 0.91   & 0.80  &  0.69  &  0.51  &  0.71  \\ [-1.0ex]
\raisebox{1.5ex}[0pt]{{\it D. melanogaster}}
 & (0.74, 0.90) & (0.72, 0.81) &   (-0.16, 0.72) & (0.65, ~0.15) &
 (-0.42, -0.15) &  (0.28, 0.59) \\
&   0.94  &  0.86   & 0.82  & 0.68  &  0.60  &  0.80  \\ [-1.0ex]
\raisebox{1.5ex}[0pt]{{\it H. pylori}}
 &(0.80, 0.91) & (0.46, 0.72) &  ( 0.06, 0.70)
&  (0.42, -0.03) & (-0.15, -0.16) &  (0.57, 0.67)  \\ 
&  0.73 &  0.52  &  0.20 &  0.37  &   0.39  &   0.10 \\ [-1.0ex]
\raisebox{1.5ex}[0pt]{{\it H. sapiens}}
 &  (0.75, 0.69) & (0.56, 0.52) &  (-0.31, 0.18) &  (0.13, ~0.37) &
 (~0.19, ~0.38) &  (-0.12, 0.02)  \\ \hline 
& 0.82  &  0.72  & 0.53  &  0.47 & 0.35 &  0.52  \\ [-1.0ex]
\raisebox{1.5ex}[0pt]{Average}
 & (0.78, 0.79) & (0.57, 0.67) & (-0.21, 0.46)
& (0.37, 0.04) &  (-0.12, -0.03) &(0.34, 0.48) \\ 
\end{tabular} 
\end{ruledtabular}
\end{table*}
%%%%%%%%%%%%%%%%%%%%%%%%%%%%%%%%%%%%%%%%%%%%%%%%%%%%%%%%%%%%
\clearpage
%%%%%%%%%%%%%%%%%%%%%%%%%%%%%%%%%%%%%%%%%%%%%%%%%%%%%%%%%%%%%%
\begin{table*}
\begin{ruledtabular}
\caption{The correlation coefficients and partial correlation between
  all possible pairs of centrality measures estimated for the random
  network of different number of vertices, N=1000, 3000, and 6000. For
  all cases, each vertex has the same average degree $\langle k
  \rangle =10$. $X$ stands for the degree centrality; while $Y$, $Z$,
  and $W$ stand for the betweenness, the eigenvector, and the
  closeness centrality, respectively. The notation for the
  partial correlation is abbreviated as Table I.} 
\begin{tabular}{l c c c c c c}
 & $R_{XY}$  & $R_{XZ}$ & $R_{YZ}$ & $R_{XW}$  & $R_{YW}$ & $R_{ZW}$  \\ 
\raisebox{1.5ex}[0pt]{N}  
  & ($R_{\bullet Z}, R_{\bullet W}$) & ($R_{\bullet Y}, R_{\bullet
  W}$)  & ($R_{\bullet Y}, R_{\bullet Z}$) & ($R_{\bullet X},
  R_{\bullet Z}$) & ($R_{\bullet X}, R_{\bullet Y}$) & ($R_{\bullet
  X}, R_{\bullet W}$) \\  \hline
     &  0.97  &  0.95 &  0.94 & 0.92  &  0.90  &  0.97 \\
\raisebox{1.5ex}[0pt]{1000}
     & (0.76, 0.86)  & (0.39, 0.61) &  (0.27, 0.69) & (0.43, -0.07) 
     & ( 0.05, -0.25) & (0.81, 0.86) \\ [1.0ex]
     &  0.98 &  0.95  &  0.94  &  0.93  &  0.88 &  0.96  \\
\raisebox{1.5ex}[0pt]{3000}
     & (0.82, 0.93) & (0.42, 0.56) & (0.14, 0.72) & (0.72, ~0.21) 
     & (-0.43, -0.23) & (0.67, 0.82) \\ [1.0ex]
     &  0.98  &  0.95 &  0.94 &  0.95 &  0.90  &  0.97   \\
\raisebox{1.5ex}[0pt]{6000}
     & (0.79, 0.89) & (0.45, 0.35) & (0.16, 0.60) & (0.77, ~0.41) &
  (-0.43, -0.10) &  (0.69, 0.83) \\
\end{tabular} 
\end{ruledtabular}
\end{table*}
%%%%%%%%%%%%%%%%%%%%%%%%%%%%%%%%%%%%%%%%%%%%%%%%%%%%%%%%%%%%
%%%%%%%%%%%%%%%%%%%%%%%%%%%%%%%%%%%%%%%%%%%%%%%%%%%%%%%%%%%%
\begin{figure}
\centerline{
\includegraphics{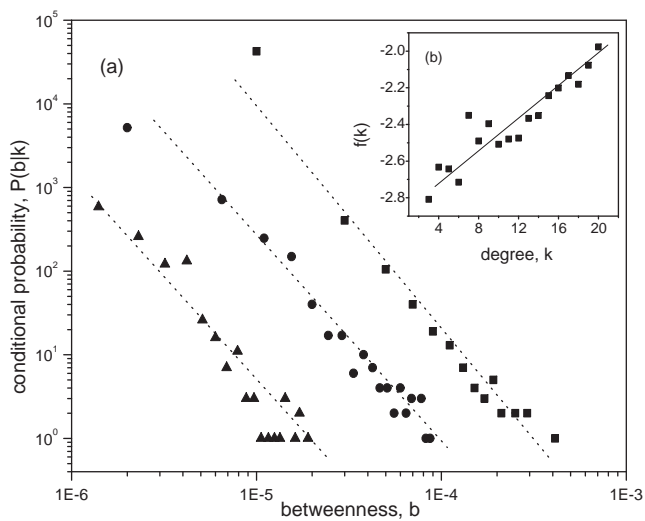}
}
\caption{(a) Log-log plots of the conditional distributions
  $P(b|k)$ of the betweenness $b$ given the degree $k$ in the film
  actor network: $k=3$ ($\blacksquare$), $k=7$ ($\bullet$), and $k=10$
  ($\blacktriangle$). The least-square fits (dotted lines) on
  the slope of $k=3$, $k=7$, and $k=10$ yield $-2.63\pm 0.12$,
  $-2.35\pm 0.10$, and $-2.51\pm 0.18$, respectively. Plots for $k=7$ and
  10 are shifted to the left for the display purpose. (b) (inset) The
  plot of the exponent $f(k)$ in Eq.~(\ref{power}) versus the degree
  $k$. Estimated values from the least square fit for
  Eq.~(\ref{linear}) are $\alpha=0.04\pm 0.01$ and
  $\beta=-2.85\pm0.05$. The errors associated with the fit are
  statistical uncertainties based on fitting a straight line.} 
\end{figure}
%%%%%%%%%%%%%%%%%%%%%%%%%%%%%%%%%%%%%%%%%%%%%%%%%%%%%%%%%%%%%%
%%%%%%%%%%%%%%%%%%%%%%%%%%%%%%%%%%%%%%%%%%%%%%%%%%%%%%%%%%%%
\begin{figure}
\centerline{
\includegraphics{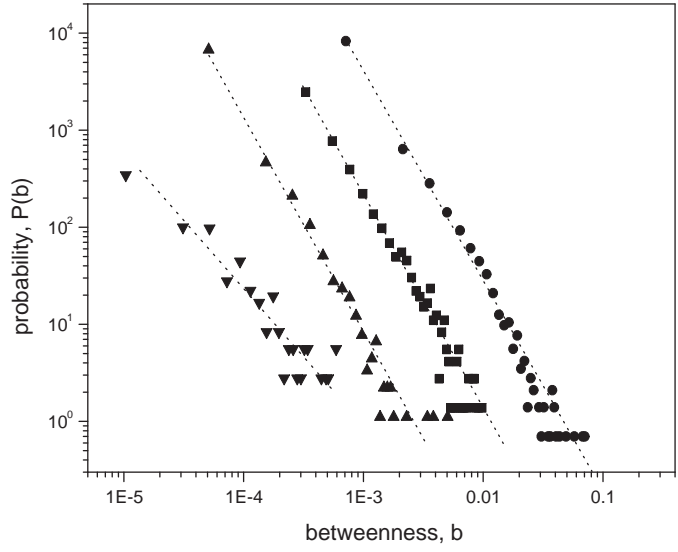}
}
\caption{Log-log plots of betweenness distributions for selected
  complex networks: the film actor network ($\blacksquare$), the
  collaboration network ($\bullet$), the protein interaction network of {\it
  D. melanogaster} ($\blacktriangle$), and the neural network of {\it
  C. elegans} ($\blacktriangledown$). Estimated exponents (dotted lines), by
  least square fits on slopes, are $2.36\pm 0.10$, $2.27\pm 0.08$,
  $2.11\pm 0.12$, and $1.31\pm 0.11$, respectively. Plots, except for
  the film actor network, are shifted horizontally for the display purpose.}
\end{figure}
%%%%%%%%%%%%%%%%%%%%%%%%%%%%%%%%%%%%%%%%%%%%%%%%%%%%%%%%%%%%%%
%%%%%%%%%%%%%%%%%%%%%%%%%%%%%%%%%%%%%%%%%%%%%%%%%%%%%%%%%%%%
\begin{figure}
\centerline{
\includegraphics{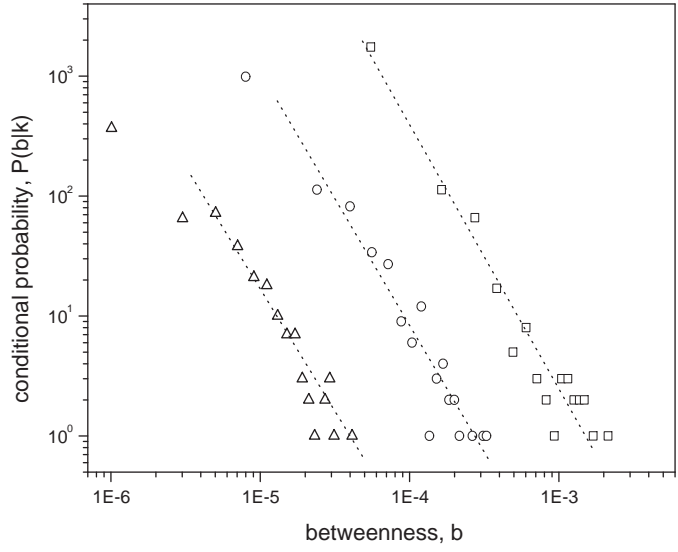}
}
\caption{Log-log plots of the conditional distributions $P(b|k)$
  of the betweenness $b$ given degree $k$ of the scientific
  collaboration network: $k=4$ ($\Box$), $k=6$ ($\bigcirc$), and $k=9$
  ($\triangle$).  The least-square fits (dotted 
  lines) on the slope of the $k=4$, $k=6$, and $k=9$ yield $-2.07\pm 0.15$,
  $-2.01\pm 0.13$, and $-2.22\pm 0.19$, respectively. Plots for $k=6$ and
  9 are shifted to the left for the display purpose.}
\end{figure}
%%%%%%%%%%%%%%%%%%%%%%%%%%%%%%%%%%%%%%%%%%%%%%%%%%%%%%%%%%%%%%
%%%%%%%%%%%%%%%%%%%%%%%%%%%%%%%%%%%%%%%%%%%%%%%%%%%%%%%%%%%%
\begin{figure}
\centerline{
\includegraphics{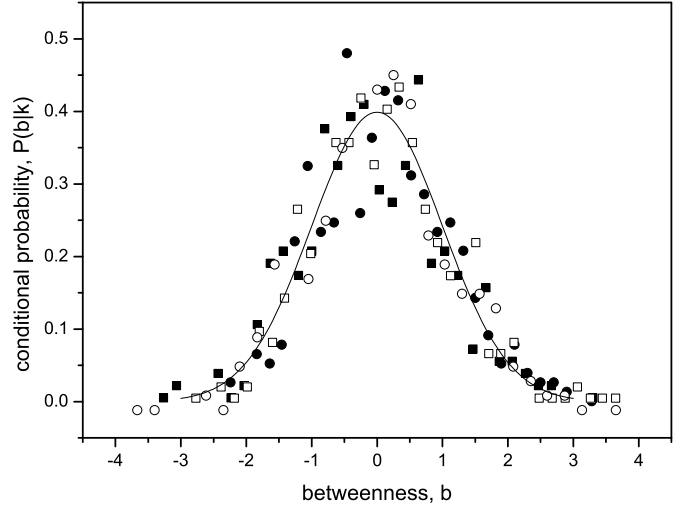}
}
\caption{Plots of the conditional distributions
  $P(b|k)$ of the betweenness $b$ given the degree $k$: $k=8$
  ($\blacksquare$), $k=10$ ($\bullet$), $k=12$ ($\Box$), and $k=14$
  ($\bigcirc$) for the random network of 3000 vertices and the average degree
  $\langle k \rangle =10$. Each distribution of different $k$ is
  normalized such that $b \rightarrow (b-{\bar b})/\sigma_{b}$, where
  ${\bar b}$ and $\sigma_{b}$ are the mean and standard deviation of
  $b$. By the normalization, all distributions collapse to the
  standard Gaussian distribution (solid line).}
\end{figure}
%%%%%%%%%%%%%%%%%%%%%%%%%%%%%%%%%%%%%%%%%%%%%%%%%%%%%%%%%%%%%%
%%%%%%%%%%%%%%%%%%%%%%%%%%%%%%%%%%%%%%%%%%%%%%%%%%%%%%%%%%%%
\begin{figure}
\centerline{
\includegraphics{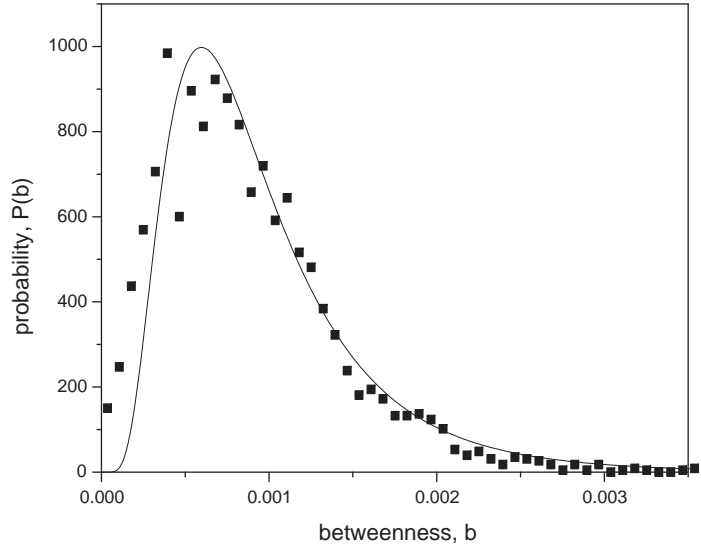}
}
\caption{The distribution of the betweenness for the
  random network of 3000 vertices and the average degree $\langle k \rangle
  =10$, together with the corresponding log-normal fit (solid line). The scale
  and shape parameters of the log-normal fit are estimated using
  the maximum likelihood estimate from the data. Estimated the scale
  and shape parameters are ${\hat \mu}=e^{-0.71}$ and ${\hat
  \sigma}=0.57$, respectively.}
\end{figure}
%%%%%%%%%%%%%%%%%%%%%%%%%%%%%%%%%%%%%%%%%%%%%%%%%%%%%%%%%%%%%%
\end{document}